\documentclass{ws-procs975x65}

\def\d{\textrm{d}}

\def\expf#1{\ensuremath{\mathrm{e}^{#1}}}
\def\tfrac#1#2{\textstyle\frac{#1}{#2}}

\def\lpderop#1{\ensuremath{\partial/\partial #1}}
\def\oder#1#2{\frac{\d #1}{\d #2}}

\begin{document}

\title{TRAPPING OF NEUTRINOS IN EXTREMELY COMPACT
  STARS\footnote{This research has been supported by Czech grants MSM 
    4781305903 and LC 06014 (M.U.).}} 

\author{ZDEN\v{E}K STUCHL\'IK, MARTIN URBANEC, GABRIEL T\"OR\"OK,
  STANISLAV HLED\'IK \\ AND JAN HLAD\'IK}


\address{Institute of Physics, Faculty of Philosophy and Science,
  Silesian University in Opava,  
Bezru\v{c}ovo n\'am. 13, Opava 746 01, Czech Republic \\
\email{zdenek.stuchlik@fpf.slu.cz}}


\begin{abstract}
Trapping of neutrinos in extremely compact stars containing trapped null
geodesics is studied. We calculated the ratio of produced to trapped neutrinos
in the simplest model of uniform density stars. This gives
the upper limit on trapping coefficients in real objects. 
\end{abstract}

\bodymatter

\section{Introduction}

Trapped null geodesics are concentrated around the stable circular geodesic
\cite{Stu-etal:2001:PHYSR4:,Abr-Mil-Stu:1993:PHYSR4:}. We suppose that
neutrinos have zero mass and can move along null geodesics in the
situations when the neutrino mean free path exceeds the star radius
$R$. Considering the internal Schwarzschild spacetimes, we use the
effective potential related to the impact parameter to calculate the
ratios of trapped to produced neutrinos.

\section{Internal Schwarzschild spacetime, effective potential and
  impact parameter}
The line element of internal Schwarzschild spacetimes of uniform density
\cite{Schw:1916:SITBA:} reads
\begin{equation}
  \d s^2 = - \expf{2\Phi(r)}\,\d t^2 + \expf{2\Psi(r)}\,\d r^2
    + r^2 (\d\theta^2 + \sin^2 \theta\,\d\phi^2).             \label{linel}
\end{equation}
The temporal and radial components of the metric tensor are given by the
formulae
\begin{equation}
  (-g_{tt})^{1/2}=\expf{\Phi}=\tfrac32Y_{1} - \tfrac12Y(r),\qquad
  (g_{rr})^{1/2} = \expf{\Psi} = 1/Y(r),                     \label{trcomp}
\end{equation}
and in geometric units $c=G=1$ there is
\begin{equation}
  Y(r) = \left(1-\frac{r^{2}}{a^{2}}\right)^{1/2},\qquad
  Y_{1}\equiv Y(R) = \left(1-\frac{R^{2}}{a^{2}}\right)^{1/2},\qquad 
  \frac{1}{a^{2}} =  \frac{2M}{R^{3}},         
\end{equation}
where $R$ is the radius and $M$ is the mass of the object. The
parameter $a$ represents the curvature of 
the internal Schwarzschild spacetime.

\begin{figure}[t]
\begin{center}
\epsfig{file=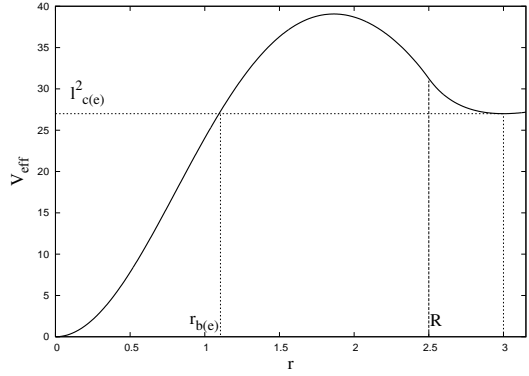,width=2.8in}
\caption{Behavior of the effective potential for
  $R=2.5GM/c^2$. Neutrinos produced in the region $r_\mathrm{b(e)}<r<R$
  are trapped if $\ell>l_\mathrm{c(e)}$ and escape when $\ell<l_\mathrm{c(e)}$.}
\label{fig01}
\end{center}
\vspace{-0.5cm}
\end{figure}

Due to the existence of two Killing vector fields, the temporal \lpderop{t}
one, and the azimuthal \lpderop{\phi} one, two conserved components of the
4-momentum must exist:
\begin{equation}
  E=-p_{t}\quad\mbox{(energy),}\qquad
  L=p_{\phi}\quad\mbox{(axial angular momentum)}.           \label{conserv}
\end{equation}
Because the motion plane is central, one can set
$\theta=\pi/2=\mathrm{const}$, choosing the equatorial plane. 
The motion along null-geodesics is independent of energy (frequency) and can
conveniently be described in terms of the impact parameter $\ell=L/E$. The
radial motion is restricted by an effective potential related to
$\ell$  by   
\begin{equation}
  \ell^{2} \leq V_{\mathrm{eff}} =\left\{
  \begin{array}{lll}
    V_{\mathrm{eff}}^{\mathrm{int}}
      = \displaystyle\frac{4a^{2}[1-Y^{2}(r)]}%
        {[3Y_{1}-Y(r)]^{2}} & \quad\mbox{for} & r\leq R\\
    V_{\mathrm{eff}}^{\mathrm{ext}}
      = \displaystyle\frac{r^{3}}{r-2M} & \quad\mbox{for} & r > R,
  \end{array}
  \right.
\end{equation}
$V_{\mathrm{eff}}^{\mathrm{int}}$ ($V_{\mathrm{eff}}^{\mathrm{ext}}$)
is the effective potential of 
the null-geodetical motion in the internal (external) Schwarzschild
spacetime~\cite{Mis-Tho-Whe:1973:Gra:}(see Fig.~\ref{fig01}).

\section{Escaped to produced neutrinos ratio}\label{ratio}

We assume that neutrinos are locally produced by isotropically emitting
sources. Then escaped-to-produced-neutrinos ratio depends on a geometrical
argument only and it is determined through directional escaping angle
$\Psi_\mathrm{e}$ given by
\begin{equation}
\sin \Psi_\mathrm{e} (r,R)= \frac{3^{3/2}}{r}\left[ \frac{3}{2}
\left(1-\frac{2M}{R}\right)^{1/2} -
\frac{1}{2}\left(1-\frac{2M}{R}\left(\frac{r}{R}\right)^{2}\right)^{1/2}\right].
\end{equation} 

\begin{figure}[t]
\begin{center}
\parbox{2.1in}{\epsfig{file=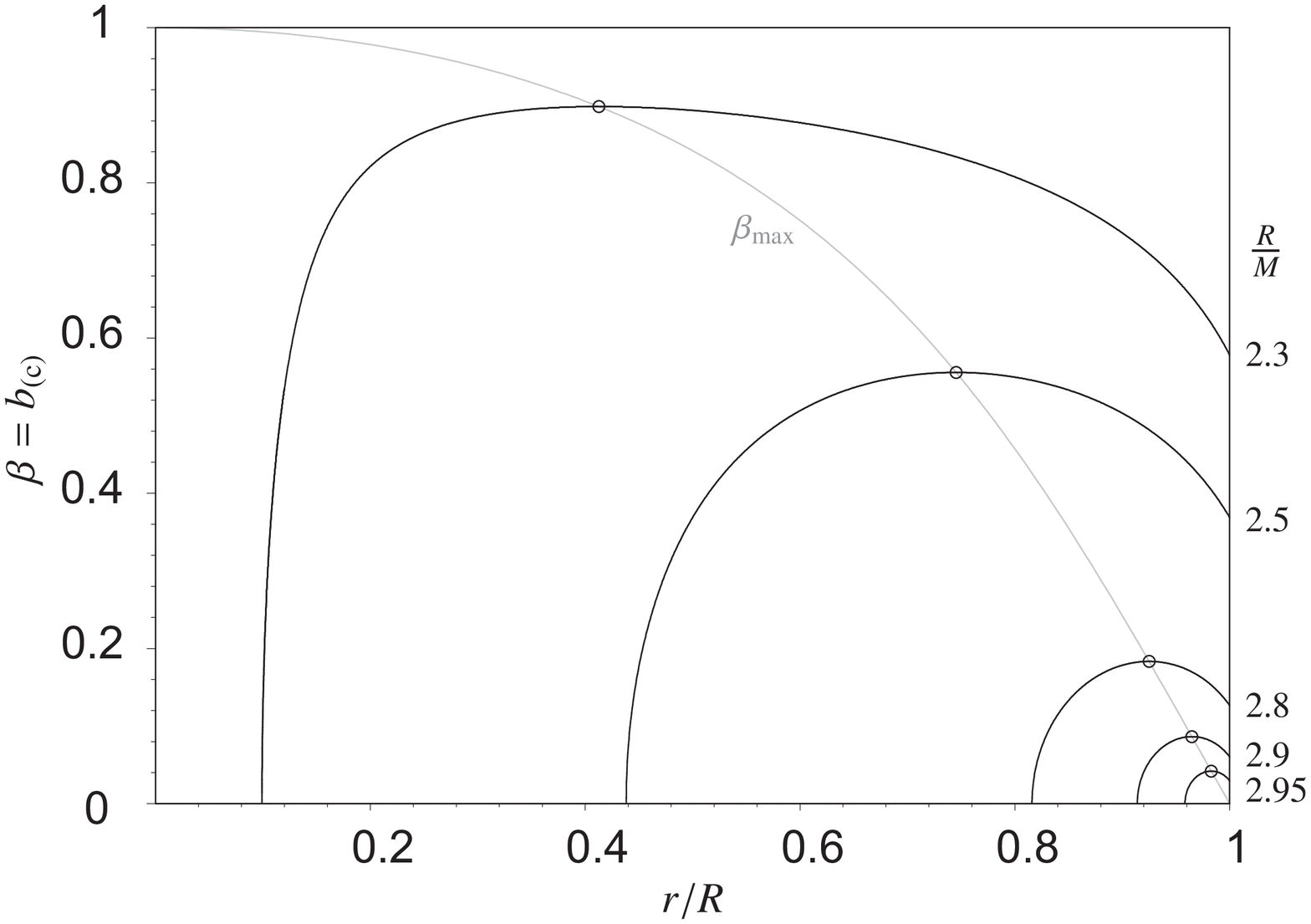,height=2.2in,angle=-90}}
\hspace*{4pt}
\parbox{2.1in}{\epsfig{file=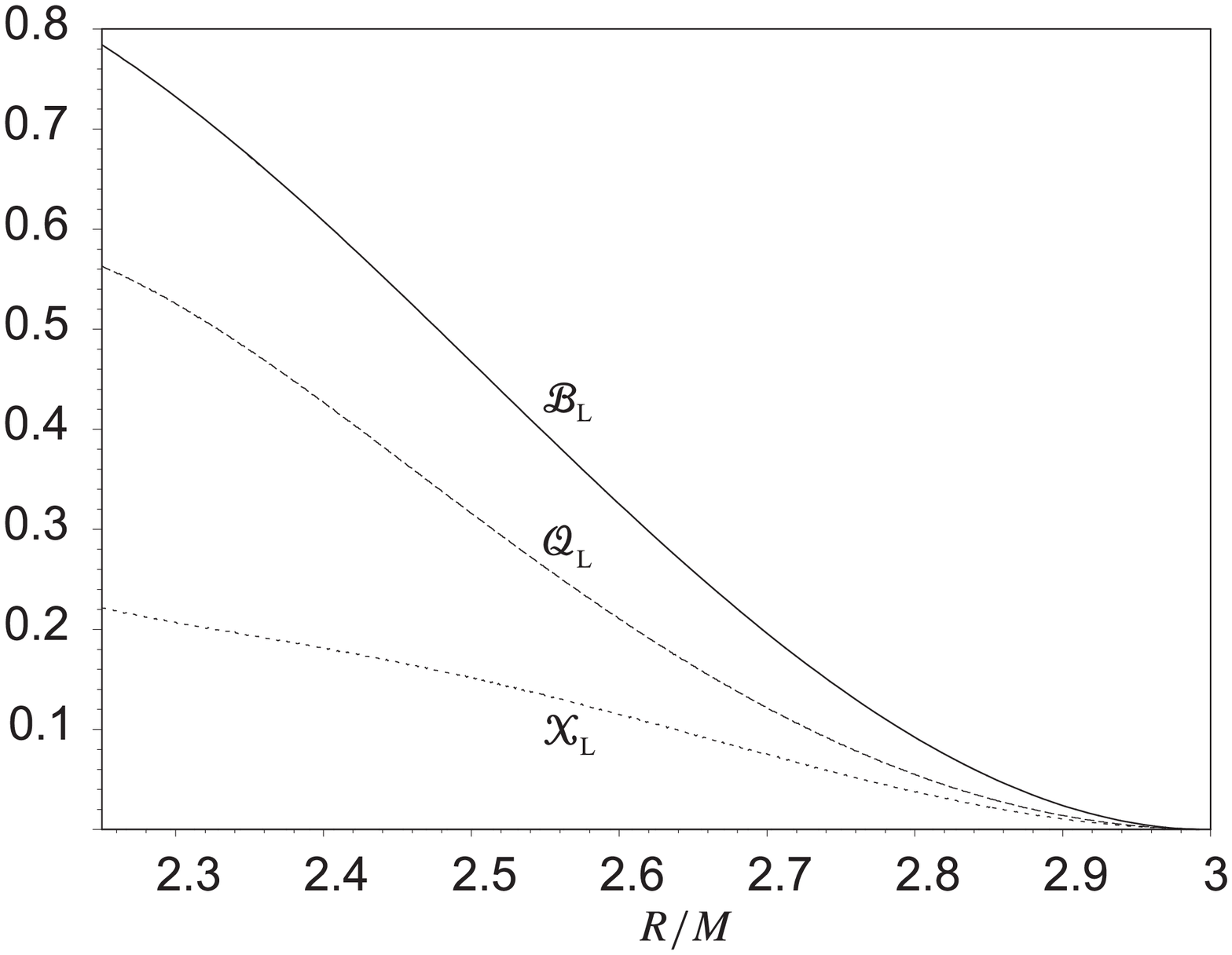,width=2in}}
\caption{Behavior of the coefficient of local trapping $\beta(r,R)$ (left)
for several values of $R/M$ and of the total coefficient of trapping
$\mathcal{B}(R)$ (right). $\mathcal{Q}_\mathrm{L}$ and
$\mathcal{X}_\mathrm{L}$ determine trapping of internal
(external)\cite{stu-etal:2004:Ragtime4and5} 
neutrinos; $\mathcal{B}_\mathrm{L}=\mathcal{Q}_\mathrm{L}+\mathcal{X}_\mathrm{L} $}
\label{fig02}
\end{center}
\vspace{-0.5cm}
\end{figure}

Let $N_{\mathrm{p}}$, $N_{\mathrm{e}}$ and $N_{\mathrm{b}}$ denote
 the number of produced, escaped and trapped neutrinos per unit
time of an external static observer at infinity. In order to determine the
global correction factors
\begin{equation}
  \mathcal{E}(R)\equiv \frac{N_{\mathrm{e}}(R)}{N_{\mathrm{p}}(R)},\qquad
  \mathcal{B}(R)\equiv \frac{N_{\mathrm{b}}(R)}{N_{\mathrm{p}}(R)}
    = 1 - \mathcal{E}(R),
\end{equation}
it is necessary to introduce the local correction factor for escaping
neutrinos at a given radius $r \in (r_{\mathrm{b(e)}},R)$. The
escaping solid angle is given by  

\begin{equation}  
\Omega_{\mathrm e}(\Psi_{\mathrm
  e})=\int\limits_0^{\Psi_{\mathrm e}}\int\limits_0^{2\pi}\sin\Psi \d\Psi\d
\phi 
  = 2 \pi (1-\cos\Psi_{\mathrm e})
\end{equation}
and the escaping (trapping) correction factors
$\epsilon(r,R)$ ($\beta(r,R)$) are given by
\begin{equation}
  \epsilon(r,R) = 1-\beta(r,R) = \oder{N_{\mathrm{e}}(r)}{N_{\mathrm{p}}(r)}
    = \frac{2\Omega(\psi_{\mathrm{e}}(r,R))}{4\pi}
    = 1 - \cos\psi_{\mathrm{e}}(r,R).
\end{equation}
The coefficient $\beta(r,R)$ determines local efficiency of the neutrino
trapping (see Fig.~\ref{fig02} for behavior of local (left) and global
(right) trapping coefficients). The global escaping (trapping)
coefficients are given by integrating the local
production rate and escaping (trapping) coefficients through the star,
and its trapping zone,
respectively\cite{stu-etal:2004:Ragtime4and5}. We assume local
production rate uniformly distributed through the star from the point
of view of local observers but it is not uniform as viewed by distant
static observers\cite{stu-etal:2004:Ragtime4and5}.

\section{Conclusions}
We have shown in the simplest case of uniform density stars that
trapping of neutrinos plays a significant role in
cooling mechanism. The local effect given by $\beta(r,R)$ is $~10\%$
for $R/M=2.95$ while the global effect described by $\mathcal{B(R)}$
is $~10\%$ for $R/M=2.87$. We expect this trapping efficiency could be
reached in some real stars with specific but realistic equations of state.


\begin{thebibliography}{00}

\bibitem{Schw:1916:SITBA:}
K.~Schwarzschild.
\newblock {\em Sitzber. Deut. Akad. Wiss. Berlin, Kl. Math.-Phys. Tech.}, pages
  424--434, 1916.

\bibitem{Stu-etal:2001:PHYSR4:}
Z.~Stuchlí\'{\i}k, S.~Hled\'{\i}k, J.~\v{S}olt\'{e}s, and E.~{\O}stgaard.
\newblock {\em Phys. Rev. D}, 64(4):044004 (17~pages), 2001.

\bibitem{Abr-Mil-Stu:1993:PHYSR4:}
M.~A. Abramowicz, J.~Miller, and Z.~Stuchl\'{\i}k.
\newblock {\em Phys. Rev. D}, 47(4):1440--1447, 1993.

\bibitem{Mis-Tho-Whe:1973:Gra:}
C.~W. Misner, K.~S. Thorne, and J.~A. Wheeler.
\newblock {\em Gravitation}.
\newblock Freeman, San Francisco, 1973.


\bibitem{stu-etal:2004:Ragtime4and5}
Z.~Stuchl\'{\i}k, Gabriel T\"or\"ok and Stanislav Hled\'\i k.
\newblock In S.~Hled\'{\i}k and Z.~Stuchl\'{\i}k, editors, {\em Proceedings of RAGtime
  4/5: Workshops on black holes and neutron stars, Opava, 14--16/13--15 October
  2002/03}, Opava, 2004. Silesian University in Opava.









\end{thebibliography}
\end{document}